\documentclass[10pt,aps,prb,twocolumn,amsmath,amssymb,superscriptaddress,nobibnotes]{revtex4-1}
\usepackage{amsmath}
\usepackage{graphicx}
\usepackage{color}
\usepackage{multirow}
\usepackage[caption = false]{subfig}
\usepackage{array}
\usepackage{threeparttable}
\usepackage[version=4]{mhchem}
\usepackage{mathtools}
\usepackage{siunitx}
\usepackage{centernot}
\usepackage{tikz}
\tikzset{font={\fontsize{11pt}{12}\selectfont}}
\usetikzlibrary{shapes,arrows}
\usepackage{lipsum}
\usepackage[normalem]{ulem}
\usepackage{hyperref}
\usepackage[capitalise,nameinlink]{cleveref}
\makeatletter
\renewcommand*{\thesection}{\arabic{section}}
\renewcommand*{\thesubsection}{\thesection.\arabic{subsection}}
\renewcommand*{\p@subsection}{}

\renewcommand*{\p@subsubsection}{}
\makeatother
\usepackage{soul}

\usepackage{hyperref}
\hypersetup{
	linkcolor=blue,
	citecolor=blue,
	urlcolor=blue,
}

\renewcommand{\thetable}{\arabic{table}}

\usepackage{lipsum}
\usepackage{orcidlink}

\begin{document}
\title{Chemical Origins of Non-Bonded Interactions Within and Between Solids} 
\author{Paul J. Robinson\orcidlink{0000-0003-0465-4979}}
\affiliation{Department of Chemistry and Chemical Biology, Harvard University, Cambridge, Massachusetts 02138, United States}
\author{Adam Rettig\orcidlink{0000-0002-6528-9576}}
\affiliation{Department of Chemistry and Chemical Biology, Harvard University, Cambridge, Massachusetts 02138, United States}
\author{Hieu Q. Dinh\orcidlink{0000-0003-1216-1234}}
\affiliation{Department of Chemistry and Chemical Biology, Harvard University, Cambridge, Massachusetts 02138, United States}
\author{Anton Z. Ni\orcidlink{0000-0003-0122-9168}}
\affiliation{Department of Chemistry and Chemical Biology, Harvard University, Cambridge, Massachusetts 02138, United States}
\author{Joonho Lee\orcidlink{0000-0002-9667-1081}}\thanks{email: joonholee@g.harvard.edu}
\affiliation{Department of Chemistry and Chemical Biology, Harvard University, Cambridge, Massachusetts 02138, United States}

\begin{abstract}
Non-bonded interactions govern structure, stability, and function across a wide range of solid-state materials, yet their chemical origins are often difficult to resolve from total energies alone. Here we generalize absolutely localized molecular orbital energy decomposition analysis to quantify and interpret non-bonded interactions within and between solids at the density functional theory level. Across molecular crystals, moir{\'e} heterobilayers, and layered perovskite heterostructures, this framework separates lattice-formation energies, interlayer binding energies, and band-structure changes into chemically intuitive contributions from frozen interactions, polarization, and charge transfer. The analysis reveals how dispersion controls polymorph stability in pharmaceutical crystals, how local stacking modulates interlayer coupling in \ce{MoS2}/\ce{WSe2}, and how alkali-cation substitution switches the quantum-well character of layered perovskite heterostructures. By connecting emergent solid-state properties to microscopic interaction mechanisms, this framework provides a chemically transparent basis for understanding and designing complex materials.\end{abstract}

\maketitle

{\it Introduction.} 
{Non-bonded interactions have become a central pillar of modern materials research.}
In molecular crystals, they determine how molecules pack and why polymorphs differ in stability. In layered materials, they control how distinct solid-state subsystems bind and electronically influence one another. These systems include molecular crystals as well as layered heterostructures such as moir{\'e} materials\cite{mak_semiconductor_2022, huang_excitons_2022} and layered perovskites,\cite{novoselov_2d_2016, aubrey_directed_2021} where weak interactions can have outsized consequences for structure and electronic properties.

Non-bonded interactions possess a rich chemical language and have motivated a large ecosystem of theoretical tools for their analysis.\cite{stone_theory_2013,
szalewicz_symmetry-adapted_2012,
patkowski_recent_2020,
mo_energy_2011,
pastorczak_perspective_2017,
mao_intermolecular_2021}
Among these, energy decomposition analysis (EDA) is especially appealing because it connects total interaction energies between subsystems, such as molecules, layers, or monomers, to chemically intuitive components, including electrostatics, Pauli repulsion, dispersion, polarization, and charge transfer. Molecular EDA has a long history of development and application,\cite{morokuma_molecular_1971, kitaura_new_1976, ziegler_calculation_1977, stoll_use_1980, gianinetti_modification_1996, mo_theoretical_1998, mo_energy_2000, khaliullin_unravelling_2007, horn_unrestricted_2013, horn_polarization_2015} and several important developments have extended decomposition ideas to periodic systems.\cite{philipsen_role_2006,podeszwa_predicting_2008,imamura_extension_2010,kuhne_electronic_2013,raupach_periodic_2015,staub_energy_2019,kuhne_cp2k_2020,niedzielski_density-based_2025} Yet chemically transparent EDA frameworks for non-bonded interactions within and between solids remain far less developed than their molecular counterparts.

Here, we generalize absolutely localized molecular orbital energy decomposition analysis (ALMO-EDA)
\cite{khaliullin_unravelling_2007, mao_intermolecular_2021,
kuhne_electronic_2013, 
staub_energy_2019}
to periodic systems, enabling microscopic analysis of non-bonded interactions within and between materials at the density functional theory level. The resulting framework can be applied to cohesive energies in molecular solids, interlayer binding energies in layered materials, and even band structures and band gaps. We use it to uncover the chemical origins of crystal-engineering trends in molecular crystals, local electronic structure changes across the moir{\'e} landscape of MoS$_2$/WSe$_2$, and the switching of quantum-well character in layered perovskite heterostructures. More broadly, this approach provides a chemically transparent framework for understanding and designing complex solid-state materials.

{\it Theory.}
We decompose the interaction energy between two subsystems $A$ and $B$ as
\begin{align}
    E^{AB}_\text{int} &= E^{AB} - E^A - E^B \nonumber \\
    &\equiv E_\text{frz} + E_\text{pol} + E_\text{ct},
    \label{eq_eda_def}
\end{align}
where $E_\text{frz}$ is the frozen interaction energy (Frz), $E_\text{pol}$ is the polarization energy (Pol), and $E_\text{ct}$ is the charge-transfer energy (CT). The frozen term contains classical electrostatics, Pauli repulsion, and dispersion. The polarization term captures \textit{intra}-subsystem electronic relaxation, whereas the charge-transfer term captures \textit{inter}-subsystem electronic relaxation. Although several quantum-chemical definitions of these terms are possible, we adopt ALMO-EDA~\cite{khaliullin_unravelling_2007, mao_intermolecular_2021} because of its variational character and its compatibility with correlated electronic-structure methods\cite{thirman_energy_2015, wang_second-generation_2025} beyond density functional theory (DFT). 
We focus here on DFT-based ALMO-EDA because DFT remains the most practical framework for solid-state electronic-structure calculations.

\begin{figure}[t]
    \centering
    \includegraphics[width=1\linewidth]{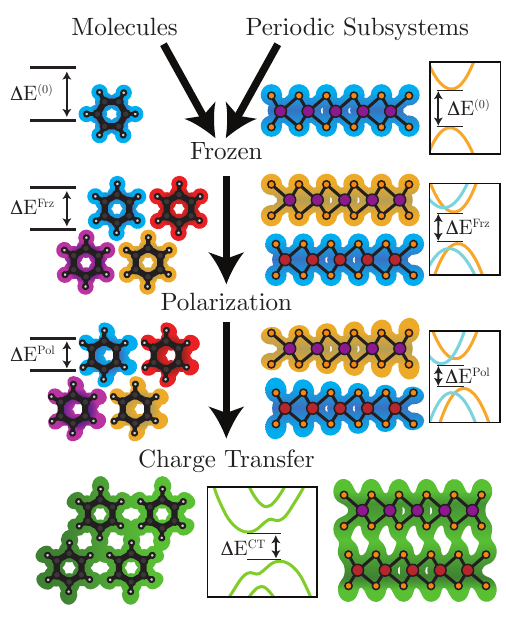}
    \caption{\textbf{Flowchart of the solid-state ALMO-EDA family of methods.}
    On the left, the scheme preserves the isolated character of each molecule until the CT step, whereas on the right, it allows charge to redistribute between unit cells within the same subsystem at every step of the EDA. 
    }
    \label{fig:infographic}
\end{figure}

For solids, ALMO-EDA operates in two distinct settings (\cref{fig:infographic}): one in which molecular subsystems are periodically repeated throughout space, and another in which extended solid-state subsystems interact with one another. As in molecular ALMO-EDA, constraints are imposed at successive stages of the electronic-structure calculation to suppress or permit charge relaxation between subsystems, thereby defining the Frz, Pol, and CT contributions. For molecular solids, we define individual molecules as subsystems and decompose cohesive or lattice energies into intermolecular interactions. 
For layered and other extended materials, we instead define extended subsystems and use the same framework to analyze interlayer interaction energies and band-structure changes. 
Further theoretical details are provided in Sec.~\ref{si_sec:theory}.

Several EDA approaches for periodic systems have been developed previously, particularly for adsorbate-surface interactions,\cite{philipsen_role_2006,imamura_extension_2010,raupach_periodic_2015,staub_energy_2019,kuhne_cp2k_2020} with some extensions to bulk interactions as well.\cite{podeszwa_predicting_2008,kuhne_electronic_2013,niedzielski_density-based_2025} Our contribution is to extend variational ALMO-EDA to directly characterize cohesive energies, interlayer binding energies, band structures, and band gaps within a unified solid-state framework. We next use this framework to resolve the chemical origins of non-bonded interactions across molecular crystals, moir{\'e} heterobilayers, and layered perovskite heterostructures.

\begin{figure*}[th]
    \centering
    \includegraphics[width=1\linewidth]{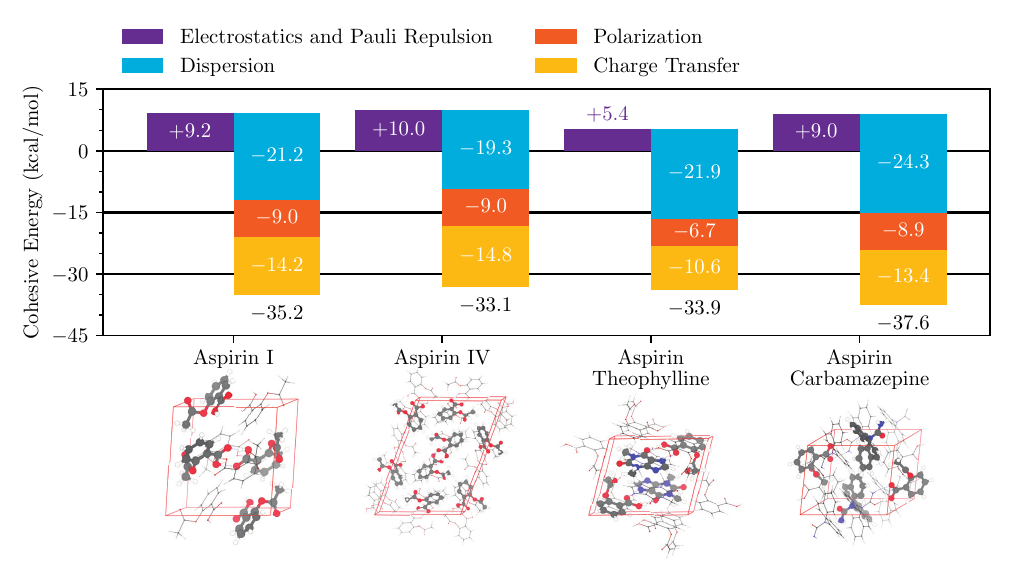}
    \caption{\textbf{ALMO-EDA applied to molecular crystals.}
    The diagrams below show the unit cell (red lines) with the subsystems shown by large atoms and the periodic repeats of the subsystems shown by small atoms.
    }
    \label{fig:molecular_crystals}
\end{figure*}

{\it Molecular Crystals.}
Pharmaceutical co-crystals are periodic solids formed from two or more molecular components and are widely studied because they can modify drug-delivery properties without changing the underlying active pharmaceutical ingredient.\cite{bolla_crystal_2022, xu_nucleic-acid-base_2024, yaghoobi_nia_co-crystal_2025} In ~\cref{fig:molecular_crystals}, we analyze two polymorphs of aspirin (I and IV) together with two aspirin-containing co-crystals, enabling direct comparison between aspirin I and 
{related solids containing active pharmaceutical ingredients.}

Aspirin I and aspirin IV are polymorphs with identical chemical composition but different molecular packing. Both structures contain similar aspirin dimers arranged to maximize hydrogen bonding between carboxylic acid groups. Because hydrogen bonds typically contain comparable Pol and CT contributions,\cite{cobar_examination_2012, horn_probing_2016} the small differences of $0.0$ kcal/mol in polarization and $0.6$ kcal/mol in charge transfer indicate that this local motif is largely conserved between the two polymorphs. By contrast, the dispersion contribution is $1.9$ kcal/mol more stabilizing in aspirin I, while electrostatics and Pauli repulsion are $0.8$ kcal/mol less favorable in aspirin IV. Thus, aspirin I is more stable because its packing yields a more favorable dispersion without substantially altering the local hydrogen-bonding interactions.

Relative to aspirin I, aspirin-theophylline has a similar dispersion contribution, differing by only $0.7$ kcal/mol, but a substantially more favorable electrostatics-plus-Pauli term, lower by $3.8$ kcal/mol. The key difference instead lies in the hydrogen-bonded aspirin-theophylline dimer, which weakens the polarization and charge-transfer contributions. Gas-phase dimer calculations support this interpretation: the $2.3$ and $3.6$ kcal/mol increases in the solid-state polarization and charge-transfer energies relative to aspirin I are mirrored by corresponding increases of {$2.1$} and $3.6$ kcal/mol in the dimers (Sec.~\ref{si_sec:molecular_eda}). These results indicate that the hydrogen bond formed between aspirin and theophylline is weaker than the carboxylic-acid dimer interaction in aspirin I. Overall, the more favorable steric environment in aspirin-theophylline does not compensate for its weaker Pol and CT, resulting in a weaker lattice energy than in aspirin I.

Aspirin-carbamazepine shows the opposite balance. Relative to aspirin I, the electrostatics-plus-Pauli contribution differs by only $0.2$ kcal/mol, whereas the dispersion contribution is $3.1$ kcal/mol more stabilizing, likely because of the additional aromatic rings on carbamazepine. At the dimer level, however, the hydrogen bond is again weaker than in aspirin I: the Pol and CT contributions differ by {$1.4$} and $2.3$ kcal/mol, respectively. In the solid state, these deficits are largely recovered, and the Pol and CT terms differ from those of aspirin I by only $0.1$ and $0.8$ kcal/mol, respectively. This suggests that interactions beyond the central dimer, likely including chalcogen-$\pi$ and $\pi$-$\pi$ contacts, compensate for the weaker local hydrogen bond. The lower cohesive energy of aspirin-carbamazepine therefore arises from a cooperative balance between enhanced dispersion and additional stabilization beyond the dimer motif.

These results show that ALMO-EDA adds a quantitative measure of bonding similarity to established crystal-engineering principles such as preserving hydrogen-bonding motifs and geometric complementarity.\cite{fleischman_crystal_2003, tothadi_shape_2011, desiraju_crystal_2013, bolla_crystal_2022} 
More broadly, these comparisons show that polymorph and co-crystal stability are governed not only by the retention of the same dominant hydrogen-bonding motifs but also by how the surrounding crystal environment redistributes dispersion, Pol, and CT.

\begin{figure*}[htb]
    \centering
    \includegraphics[width=1\linewidth]{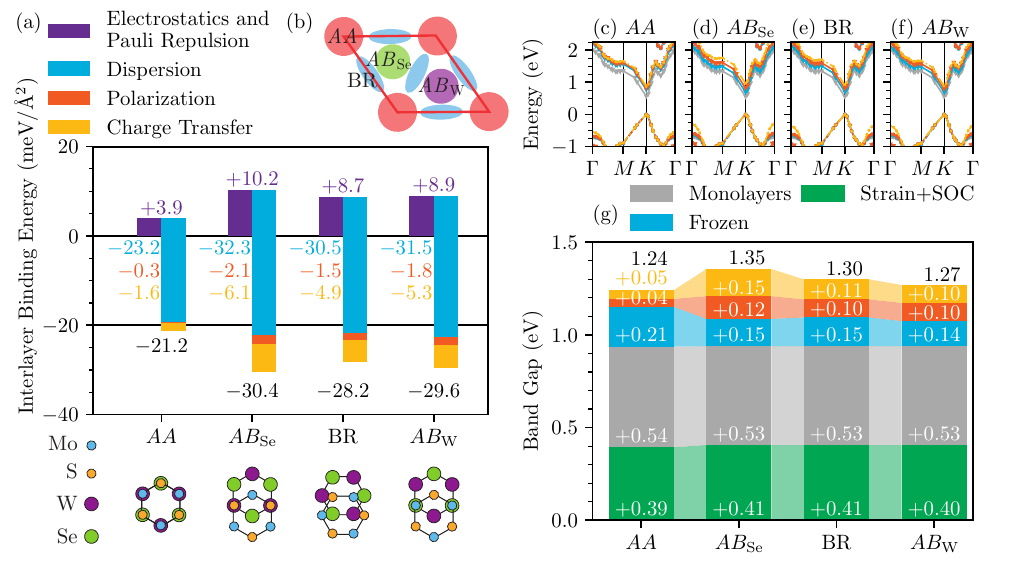}
    \caption{
    \textbf{ALMO-EDA applied to the MoS$_2$/WSe$_2$ heterobilayer.}
    (a) Interlayer binding energy across the structures in a moir{\'e} potential.
    (b) Cartoon of a moir{\'e} heterobilayer showing the connectivity between regions.
    (c-f) EDA decomposed band structures across the structures.
    (g) Band-gap decomposition, including a correction term for the spin-orbit coupling (SOC) and strain of the monolayers.
    }
    \label{fig:bilayer_eda}
\end{figure*}

{\it Twisted Bilayers.}
The \ce{MoS2}/\ce{WSe2} heterobilayer hosts a moir{\'e} exciton in which the hole resides on the WSe$_2$ layer and the electron on the MoS$_2$ layer.\cite{zhang_interlayer_2017} Here, we approximate different regions of the moir{\'e} superlattice by the corresponding untwisted local stacking arrangements, following the strategy of Ref.~\citenum{zhang_interlayer_2017}. In \cref{fig:bilayer_eda}(a), we show the ALMO-EDA of the interlayer binding energy for the four prominent stacking domains illustrated in \cref{fig:bilayer_eda}(b).

As expected for a van der Waals heterostructure, the interlayer binding energetics are dominated by the Frz contribution, especially dispersion. The dispersion contribution is approximately $-30$ meV/\AA$^2$ for the three more stable stackings, whereas the much less stable $AA$ domain has a substantially smaller value of $-23.2$ meV/\AA$^2$. Among the more stable domains, the dispersion contribution follows the overall stability trend: relative to $AB_\text{W}$, it is $1.0$ meV/\AA$^2$ less stabilizing in BR and $0.8$ meV/\AA$^2$ more stabilizing in $AB_\text{Se}$.

The electrostatics-plus-Pauli contribution increases as the interlayer separation decreases. Using the transition-metal $z$ coordinates to define the layer spacing, the interlayer distance is $7.02$ \AA\ in $AA$, compared with $6.54$ \AA\ in BR, $6.48$ \AA\ in $AB_\text{W}$, and $6.43$ \AA\ in $AB_\text{Se}$. Consistent with this trend, the electrostatics-plus-Pauli contribution rises from $3.9$ meV/\AA$^2$ in $AA$ to $8.7$ meV/\AA$^2$ in BR, $8.9$ meV/\AA$^2$ in $AB_\text{W}$, and $10.2$ meV/\AA$^2$ in $AB_\text{Se}$. The Pol and CT contributions follow the same qualitative trend: both are smallest in $AA$, at $-0.3$ and $-1.6$ meV/\AA$^2$, respectively, and become increasingly stabilizing as the layers approach one another. Overall, the stacking energetics are well described by Frz interactions, but $AA$ remains qualitatively distinct from the other domains in its much smaller Pol and CT contributions.

This distinction becomes much more important in the band-gap decomposition. Although dispersion dominates the interlayer binding energy, it does not contribute to the band structure at the level of theory used here. Because lattice mismatch between \ce{MoS2} and \ce{WSe2} generates a moir{\'e} superlattice, different local stacking domains produce different local band gaps and thus an electronic superlattice. In \cref{fig:bilayer_eda}(c-f), we show the band-structure evolution across the EDA steps and find that each stage increases the band gap while preserving the overall band character, including the $K \rightarrow K$ direct gap. The isolated-monolayer contribution is approximately $0.53$ eV in all four domains and corresponds to the excitation energy of two infinitely separated layers. As discussed in Sec.~\ref{si_sec:MoS2WSe2_SOC_and_strain}, we additionally include corrections for spin-orbit coupling (SOC) and artificial lattice strain: SOC reduces the gap by about $0.2$ eV, whereas strain increases it by about $0.6$ eV.

The $AA$ domain has the smallest band gap, $1.24$ eV, and is dominated by a Frz contribution of $0.21$ eV, with much smaller dispersion and CT contributions of $0.04$ and $0.05$ eV, respectively. By contrast, $AB_\text{Se}$ has the largest band gap, $1.35$ eV, with a smaller Frz contribution of $0.15$ eV and substantially larger Pol and CT contributions of $0.12$ and $0.15$ eV, respectively. BR is intermediate in character, with Frz, Pol, and CT contributions of $0.15$, $0.10$, and $0.11$ eV, respectively, while $AB_\text{W}$ has the second smallest band gap, $1.27$ eV, with still smaller Frz and CT contributions of $0.14$ and $0.10$ eV. With the combination of hybrid DFT and the strain-plus-SOC correction, the absolute band gaps are also close to experiment. For example, the reported band gap of $AB_\text{W}$ is $1.14$ eV,\cite{zhang_interlayer_2017} and the experimentally observed differences between the $AA$ band gap and the $AB_\text{Se}$, BR, and $AB_\text{W}$ band gaps, approximately $0.17$, $0.09$, and $0.01$ eV, are reasonably reproduced by our calculated differences of $0.11$, $0.06$, and $0.03$ eV, respectively.

Taken together, these results show that moir{\'e}-domain energetics are governed largely by Frz interactions, whereas the character of the low-energy interlayer excitation depends much more strongly on Pol and CT. In particular, the smaller Pol and CT contributions in $AA$ imply that its interlayer excitation has less interlayer hybridization than the corresponding excitations in the other domains. 
This identifies Pol and CT, rather than interlayer dispersion interactions, as the chemical drivers that control how local stacking modulates interlayer excitations across the moir{\'e} landscape.

\begin{figure*}[htb]
    \centering
    \includegraphics[width=1\linewidth]{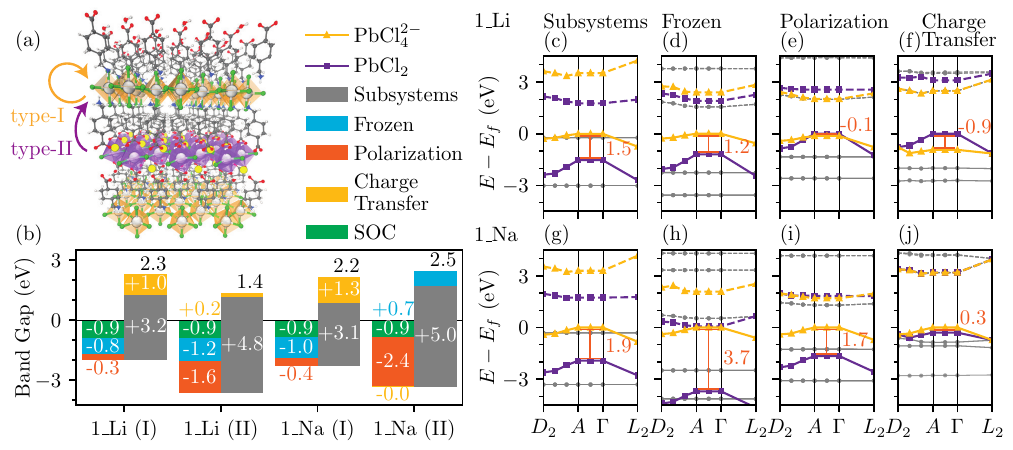}
    \caption{
    \textbf{ALMO-EDA applied to layered perovskite heterostructures.}
(a) Schematic of the competing type-I and type-II quantum wells.
(b) Band-gap decompositions for 1\_Li and 1\_Na.
(c)-(f) Band structures and densities of states for 1\_Li across the EDA steps.
(g)-(j) Band structures and densities of states for 1\_Na across the EDA steps.
In (c)-(j), {only the highest occupied (solid lines) and lowest unoccupied bands (dashed lines) are shown for each subsystem.
The gray lines denote the molecular subsystems.
The red annotated values indicate the difference between the valence band maxima of \ce{PbCl4^{2-}} and \ce{PbCl2}}. 
    }
    \label{fig:perovskite_eda}
\end{figure*}

{\it Perovskite Heterostructures.}
Recently, Deshmukh et al.\ showed that alkali-cation substitution can switch the quantum-well character of layered perovskite heterostructures.\cite{deshmukh_tuning_2025} In the Na-containing material, both the electron and hole are localized in the perovskite layer, giving a type-I quantum well, whereas in the Li analog, the electron remains in the perovskite layer but the hole localizes in the intergrowth PbCl$_2$ layer, giving a type-II quantum well. We illustrate these two possibilities in \cref{fig:perovskite_eda}(a). 
{Here, we ask how bringing together the \ce{PbCl4^{2-}} perovskite layer, 
the \ce{PbCl2} intergrowth layer, 
and the other constituent molecules, 
drives this switch in quantum-well character.
}

The heterostructures have the chemical formula M$_2$(PbCl$_2$)(AMCHC)$_2$(PbCl$_4$)$\cdot$2H$_2$O (M = Li, Na), and we follow the notation of Ref.~\citenum{deshmukh_tuning_2025} by denoting them 1\_Li and 1\_Na. 
The band-gap EDA in \cref{fig:perovskite_eda}(b) shows that 1\_Li adopts a type-II band gap because the Frz and Pol shrink the type-II gap by 1.7 eV more than the type-I gap, and CT increases the type-I gap by 0.8 eV more than the type-II gap.
Together, these overcome the 1.6 eV preference towards type-I gaps in the isolated subsystems.
By contrast, 1\_Na remains type-I because the 2.0 eV and 1.3 eV Pol and CT driven preferences towards type-II, combined with a 0.3 eV preference towards type-I of the Frz, are not large enough to overcome the 1.9 eV difference between type-II and type-I at the subsystem level.

From the band-gap decomposition, we see a pattern that Pol and CT consistently push towards type-II gaps.
We note that the contribution of the SOC to the band gaps of both 1\_Li and 1\_Na is an approximately constant -0.9 eV decrease across all gap types (See Sec.~\ref{si_ssec:perovskite_soc} for further details).
We note that DFT without exact exchange is known to significantly underestimate band gaps.\cite{perdew_density_1985}
Here, the DFT band gaps are lower than the experimental optical gap of $\approx 3.5$ eV, which is itself lower than the true band gap because of exciton formation. 
With that caveat in mind, trends in perovskite band gaps can be well-described by DFT.\cite{filip_screening_2022} Additionally, for these materials, the principal question is the ordering of the orbitals in the valence band.

The band-structures evolution reveals more nuance into the valence band ordering, which determines the type of band-gap. In particular, it highlights the role of the Frz step, which is obscured in the band-gap decomposition by simultaneous shifts in the valence and conduction bands.
In the isolated subsystems, shown in \cref{fig:perovskite_eda}(c) and \cref{fig:perovskite_eda}(g), the system shows a type-II gap, but from the perovskite valence band to the intergrowth conduction band.
The isolated subsystems, \cref{fig:perovskite_eda}(c) and \cref{fig:perovskite_eda}(g), establish a baseline for the valence band splitting between \ce{PbCl4^{2-}} and \ce{PbCl2}. 1\_Li and 1\_Na have splittings of 1.5 eV and 1.9 eV, respectively.
Advancing to the Frz step, \cref{fig:perovskite_eda}(d) and (e), the intergrowth valence band moves towards the Fermi level by 0.3 eV in 1\_Li and away from the Fermi level by 1.8 eV in 1\_Na.
This is a key indication of the chemistry, but is obscured by the band-gap decomposition because the \ce{PbCl4^{2-}} and \ce{PbCl2} conduction bands also shift downward by comparable amounts.

At the Pol step, \cref{fig:perovskite_eda}(e) and \cref{fig:perovskite_eda}(i), the intergrowth valence band shifts upward in both materials, but only in the Li compound does this shift produce a type-II quantum well. Charge transfer produces similar additional upward shifts of the \ce{PbCl2} valence band, \cref{fig:perovskite_eda}(f) and \cref{fig:perovskite_eda}(j), but does not alter the quantum-well type.
We note that because CT allows orbital mixing between subsystems, assignment of bands to specific subsystems is not uniquely defined.
In the case of \cref{fig:perovskite_eda}(j), our visualization assigns both layers to the valence band. 
In Sec.~\ref {si_ssec:perovskite_mulliken}, we present the subsystem breakdown of the bands in more detail and show that \ce{PbCl4^{2-}} is the predominant component of the 1\_Na conduction band.

The switch in quantum-well character reflects a competition between Frz interactions and electronic relaxation, and can be traced to the different ionic sizes of Li$^+$ and Na$^+$. In 1\_Na, stronger electrostatics-plus-Pauli repulsion between Na$^+$ and the PbCl$_2$ intergrowth makes hole formation on the intergrowth substantially less favorable, thereby stabilizing the type-I quantum well despite the persistent tendency of Pol and CT to favor type-II alignment. In 1\_Li, the smaller cation perturbs the intergrowth less strongly, allowing Pol and CT to stabilize the type-II state.

While the Frz contribution sets the stage for the eventual quantum-well character.
The magnitudes of Pol and CT determine whether a type-II switch occurs.
This suggests their use as an accessible handle for tuning related materials. 
More broadly, this analysis shows how a chemically simple substitution can switch the quantum-well character by altering the balance between Frz interactions and electronic relaxation at the interface.

{\it Conclusions.}
We have generalized ALMO-EDA to resolve the chemical origins of non-bonded interactions within and between solids. Across molecular crystals, moir{\'e} heterobilayers, and layered perovskite heterostructures, the resulting framework connects cohesive energies, interlayer binding energies, and electronic structure to chemically intuitive contributions from frozen interactions, polarization, and charge transfer. In doing so, it distinguishes between cases in which solid-state stability and function are primarily governed by frozen interactions and those in which polarization and charge transfer become decisive.

These applications show that non-bonded interactions in solids can be understood in the same chemically intuitive language long used for molecular systems, while also revealing design principles unique to extended materials. Looking forward, extensions to spin-orbit-coupled systems, operando electrochemical environments~\cite{ni_gaussian-based_2025}, and correlated methods beyond DFT should further broaden the scope and predictive power of this approach. Overall, this work establishes a chemically transparent framework for understanding how weak interactions shape the structure and electronic properties of solids.

{\it Data Availability.}
Q-Chem, QC-PBC, and Quantum Espresso input and output files, extracted data, and python scripts to reproduce figures and tables presented in this work are available at \url{https://github.com/JoonhoLee-Group/solid_state_almo_eda_data}.

{\it Acknowledgments.}
This work was supported by a startup fund from Harvard University.
P.J.R. acknowledges the support of the Arnold and Mabel Beckman Foundation
(\url{http://dx.doi.org/10.13039/100000997})
through the Arnold O. Beckman Postdoctoral Fellowship.
The work of A. R. was also supported by the Harvard Quantum Initiative prize postdoctoral fellowship.
The work of A. Z. N. was also supported by the NSF GRFP Fellowship.
We thank Prof. Hemamala Karunadasa for suggesting the perovskite heterostructure as an application of our ALMO-EDA.
This work used computational resources
from FASRC supported by the FAS Division of Science Research Computing Group at Harvard, and, the Delta system at the National Center for Supercomputing Applications [award OAC 2005572] through allocation  CHE250044 from the Advanced Cyberinfrastructure Coordination Ecosystem: Services \& Support (ACCESS) program, which is supported by National Science Foundation grants \#2138259, \#2138286, \#2138307, \#2137603, and \#2138296.\cite{boerner_access_2023}

\section*{Methods\label{sec:methods}}
All EDA calculations were run with a development version of QC-PBC, the periodic code associated with the Q-Chem software package.\cite{epifanovsky_software_2021, robinson_condensedphase_2025}
{
In Secs.~\ref{si_sec:noble_gas} and~\ref{si_sec:benzene} we present validation of our ALMO-EDA method 
on both an artificial neon crystal and benzene, a common benchmarking system.\cite{wen_practical_2012, yang_ab_2014, moellmann_dft-d3_2014}
}
{Sec.~\ref{si_sec:water_surface} contains a case study on water adsorption to demonstrate reasoning for our basis-set selections with ALMO-EDA.}
All geometry relaxations were performed with Quantum Espresso using PBE and PAW pseudopotentials.
\cite{giannozzi_advanced_2017, perdew_generalized_1996, blochl_projector_1994}
Visualizations of molecular structures were created with VMD.\cite{humphrey_vmd_1996}

\subsection*{Molecular Crystals}

The pharmaceutical crystals were optimized from the starting positions from the Cambridge Crystallographic Data Centre with deposition numbers: 1101021 (Aspirin I), 1541529 (Aspirin IV), 1821703 (Aspirin Theophylline) and 1877787 (Aspirin Carbamazepine).\cite{kim_structure_1985, shtukenberg_third_2017, darwish_new_2018, nicolai_crystal_2019} 

The optimization used the PAW-PBE pseudopotentials and a 60 Ry.\ energy cutoff with the following Monkhorst-Pack grids: 
$3\times2\times3$ (benzene), 
$2\times3\times2$ (aspirin I), 
$1\times4\times1$ (aspirin IV), 
$3\times2\times2$ (Aspirin Theophylline), and 
$2\times2\times2$ (Aspirin Carbamazepine).
The Grimme-D3 correction was additionally included.

For the EDA, we applied the PBE functional\cite{perdew_generalized_1996} in combination with the D3(0) dispersion correction\cite{grimme_consistent_2010} in an all-electron calculation employing the def2-TZVP basis and rij-def2-TZVP auxiliary basis.\cite{weigend_balanced_2005, weigend_ri-mp2_1998, pritchard_new_2019, dinh_efficient_2026}
We used a uniform $3\times3\times3$ Monkhorst-Pack grid across all 5 systems. 
The SCF and SCF-MI calculations were converged via GDM or L-BGFS-LS to an error tolerance of $10^{-5}$.\cite{van_voorhis_geometric_2002}
{We included the BSSE correction of Boys and Bernardi\cite{boys_calculation_1970} for each subsystem by including `ghost' atomic centers within 4~\AA \ of the active subsystem. 
We also include corrections to account for errors introduced by the density-fitting approximation.
Both the BSSE and density fitting corrections are discussed in Sec.~\ref{si_sec:df_errors}.
}

\subsection*{Twisted Bilayers}
The MoS$_2$/WSe$_2$ EDA calculations used the HSE-HJS hybrid functional combined with the TZVP-MOLOPT-PBE0-GTH basis and GTH-PBE0 pseudo potential.\cite{krukau_influence_2006, henderson_generalized_2008, vandevondele_gaussian_2007}
A $K$-centered {$7\times7\times1$} k-mesh was applied for the EDA and band-gap analysis.
{A BSSE correction\cite{boys_calculation_1970} was additionally included in the EDA.}
The band structures were generated with displacements of a
$5\times5\times1$ k-mesh.
In Sec.~\ref{si_sec:MoS2WSe2_convergence}, we demonstrate that these k-mesh sizes are sufficiently converged.

Geometry optimizations were performed with Quantum Espresso for the four stacking arrangements.
The optimizations of $AA$, $AB_\text{Se}$ and $AB_\text{W}$ allowed the lattice constant to optimize.
BR was fixed to the average lattice constant of $AB_\text{Se}$ and $AB_\text{W}$, and the atoms were only allowed to optimize along the z-axis.
The Grimme-D3(0) correction was applied, and a 
60 Ry energy cutoff was applied. An $11\times11\times1$ k-mesh was applied for all optimizations.\cite{grimme_consistent_2010}  
To avoid extraneous inter-layer interactions, the unit cells had a height of 30 \AA \ (corresponding to a vacuum layer of $\approx 20$ \AA).

\subsection*{Perovskites Heterostructures}
The perovskite heterostructure geometries were the experimental crystal structures from Ref.~\citenum{deshmukh_tuning_2025} with Cambridge Crystallographic Data Centre deposition numbers 2445552 and 2445554.
The orientation of the disorder groups was selected to maximize the similarity with the theoretical structures presented in Ref. \citenum{deshmukh_tuning_2025}.
Following the DFT protocol established in Ref. \citenum{deshmukh_tuning_2025} we applied a $6\times6\times4$ k-mesh with the PBE functional.
Dispersion was included via the D3(BJ) correction.\cite{grimme_effect_2011}
We used the TZVP-MOLOPT-PBE-GTH basis set for H, Li (3 valence electrons), C, N, O, and Na (9 valence electrons), the aug-TZVP-GTH basis set with 7 valence electrons for Cl, and the TZVPP-MOLOPT-PBE-GTH with 22 valence electrons for Pb.\cite{vandevondele_gaussian_2007}
We applied a kinetic energy cutoff of 2000 eV.

For the EDA, we partition the systems into a (PbCl$_4$)$^{2-}$ layer, 4 separate AMCHC, 4 separate  M$^+$, 4 separate H$_2$O, and the PbCl$_2$ layer. 
All individual subsystems were simulated in the periodic unit cell.
Because periodic charged fragments in isolation do not produce physical results, for the initial step of EDA (isolated subsystems) we include the local linear polarizable continuum solvation model simulating water and non-interaction ions of unit charge.\cite{ni_gaussian-based_2025}

The band gaps were computed from the valence/conduction band minimum/maximum of the $6\times6\times4$ k-mesh.
Because the band minima and maxima are not necessarily at these points, we expect this to introduce a slight error in the gaps.
The charge transfer bands were assigned to subsystems based on Mulliken population analyses of each band. 
The subsystem assignments were made based on the first valence or conduction band with at least 15\% character of the subsystem. 
{
Because the Mulliken partitioning included band crossings between subsystems, we restricted the charge transfer valence band maximum to lie on the $A-\Gamma$ path of k-space.
}
Details of this analysis are given in Sec.~\ref{si_ssec:perovskite_mulliken}.

\bibliographystyle{achemso}
\bibliography{references}
\setcounter{figure}{0} 
\renewcommand{\thefigure}{S\arabic{figure}} 
\setcounter{table}{0} 
\renewcommand{\thetable}{S\arabic{table}} 
\renewcommand{\thesection}{S\arabic{section}}
\renewcommand{\thesubsection}{\thesection.\arabic{subsection}}
\renewcommand{\appendixname}{}
\setcounter{equation}{0}
\renewcommand{\theequation}{S\arabic{equation}}

\clearpage
\onecolumngrid
\section*{Supplementary information}
\section{Formalisms of Solid-State ALMO-EDA\label{si_sec:theory}}

The fundamental constraint of ALMO-EDA
is holding the {\it occupied} molecular-orbital (MO) coefficient matrix 
to be block-diagonal in the fragments
\begin{equation}
\mathbf C_\text{occ} =
\begin{pmatrix}
\mathbf C_A & \mathbf 0 \\
\mathbf 0   & \mathbf C_B 
\end{pmatrix},
\label{eq:gamma_almo_mo}
\end{equation}
where the rows are for atomic orbitals and the columns are for occupied orbitals.
Here, for simplicity, we consider two fragments, $A$ and $B$.
Each block is
used to create 
fragment-wise molecular orbitals that are orthonormal within each fragment, but 
non-orthogonal across different fragments.

This non-orthogonality constraint allows interacting fragment orbitals to retain the same structure as isolated fragments, thereby eliminating charge transfer between them via a variational constraint.
Beyond this constraint, various versions of ALMO-EDA are distinguished by the procedure by which occupied and virtual orbitals are assigned to a specific fragment.\cite{stoll_use_1980, gianinetti_modification_1996, horn_polarization_2015}
For each choice of partitioning, constrained SCF theories, called SCF for molecular interactions (SCFMI), have been developed to minimize the energy while maintaining the ALMO constraint.
The optimized SCFMI wavefunction and energy define the polarization contribution in conjunction with the frozen energy.

In the case of $\Gamma$-point SCFMI, 
the methodology is identical to the molecular case
except that the basis is composed of Bloch atomic orbitals.
In the following three subsections, we derive the SCFMI equations for $\Gamma$-point EDA, as well as for interactions between bulk fragments, and between molecular fragments.

\subsection{Review of \texorpdfstring{$\Gamma$}{Gamma}-point ALMO-EDA\label{si_sec:gamma_eda_theory}}

The development of ALMO-EDA for periodic systems closely resembles the molecular case, and for a full treatment of that case, we refer the reader to Ref. \citenum{horn_unrestricted_2013} and related papers. 
Throughout the following discussion, we use the same contravariant-covariant tensor notation as in Ref. \citenum{horn_unrestricted_2013}. 
$X,Y, \dots$ indicate fragment indices, 
$\mu, \nu, \dots$ indicate AO indices, 
$i, j, \dots$ indicate occupied MO indices, 
and $a, b, \dots$ indicate virtual MO indices.

Expressed in tensor notation, \cref{eq:gamma_almo_mo} becomes,
\begin{equation}
\psi_{Xi}(\mathbf r)
=
\sum_{\mathbf R}
\sum_{\mu}
C^{X\mu\bullet}_{\bullet Xi}
\phi_{X\mu}^{\mathbf R}(\mathbf r)
\end{equation}
where $\mathbf R$ is the lattice summation.
We then have a covariant metric tensor between occupied orbitals due to non-orthogonality,
\begin{equation}
\sigma_{XiYj}
=
\sum_{X\mu Y\nu}
(\mathbf C^\dagger)_{Xi \bullet}^{\bullet X\mu}
S_{X\mu Y\nu}
C_{\bullet Yj}^{Y\nu\bullet},
\end{equation}
which is then used to define
one-particle reduced density matrix (1-RDM),
\begin{equation}
P^{X\mu Y\nu}
=
C_{\bullet Xi}^{X\nu\bullet}
\sigma^{XiYj}
(\mathbf C^\dagger)_{Yj \bullet}^{\bullet Y\nu}
\end{equation}
This 1-RDM is then used to build the Fock matrix as usual, and the subsequent total ALMO-SCF energy is defined as
\begin{equation}
E = 
\frac12
\text{Tr}\left[(
\mathbf h + \mathbf F) \mathbf P\right].
\label{eq:energy}
\end{equation}

The essence of SCFMI is to minimize \cref{eq:energy} with respect to $\mathbf C$ while keeping it
block diagonal in fragments. 
While we follow Stoll's projector approach in this work, the projector of Gianinetti can also be used with identical energetic results.\cite{stoll_use_1980, gianinetti_modification_1996}
{We note that while the projectors guarantee the same total energy, they do not guarantee the same orbital energies that we use for band structure analysis.
The differences induced by this choice have not been explored.}
Within the Stoll framework, the stationary condition for each
fragment $X$ is

\begin{equation}
\mathbf 0
=
\left[
\left(\mathbf I - \mathbf S \mathbf P\right)
\mathbf F
\mathbf C
\mathbf{\boldsymbol{\sigma}}^{-1}
\right]_{XX},
\end{equation}
where
the right-hand side is the fragment-wise occupied-virtual rotation (i.e., the gradient).
Using a gradient based approach such as
geometric direct minimization (GDM),\cite{van_voorhis_geometric_2002}
we can directly minimize the ALMOs.

\subsection{Solid-state ALMO-EDA for bulk fragments\label{si_sec:band_eda_theory}}

Solid-state SCFMI is needed for the cases where we have two fragments as extended systems, such as interlayer, bulk-bulk, bulk-solvent, and surface-adsorbate interactions. 
Solid-state SCFMI formalism is equivalent to the $\Gamma$-point method in a large supercell where fragments extend across the entire supercell.
In terms of the ALMO constraint, this means enforcing block-diagonal orbitals while allowing mixing of the same fragment across different unit cells.
Here, we again consider two fragments for simplicity, and a supercell with two unit cells for illustrative purposes. This naturally leads to a block diagonal momentum space ALMO constraint of
\begin{equation}
\mathbf C_\text{occ}^\text{band}(\mathbf{k}) =
\begin{pmatrix}
\mathbf C_{A}(\mathbf{k}) & \mathbf 0\\
\mathbf 0 &  C_{B}(\mathbf{k})\\
\end{pmatrix}.
\label{si_eq:band_eda_mat}
\end{equation}
For the evaluation of the frozen energy, $C_{X}(\mathbf{k})$ is determined by a periodic calculation of the isolated fragment solid.

In our usual notation \cref{si_eq:band_eda_mat} becomes
\begin{equation}
\psi_{Xi}^{\mathbf k}(\mathbf r)
=
\sum_{\mathbf R}
\sum_{\mu}
C^{X\mu\bullet}_{\bullet Xi}({\mathbf k})
\phi_{X\mu}^{\mathbf R}(\mathbf r)
\exp(i\mathbf k \mathbf R).
\end{equation}
Our covariant metric tensor is now given by 
\begin{equation}
\boldsymbol{\sigma}(\mathbf{k})
=
\left\langle 
\psi_{Xi}^{\mathbf k}
\big|
\psi_{Yj}^{\mathbf k}
\right\rangle
=
\mathbf{C}^\dagger(\mathbf{k})\mathbf{S}(\mathbf{k})\mathbf{C}(\mathbf{k}),
\end{equation}
and the SCFMI 1-RDM is given by 
\begin{equation}
    \mathbf{P}(\mathbf{k}) = 
    \mathbf{C}(\mathbf{k})
    \boldsymbol{\sigma}^{-1}(\mathbf{k})
    \mathbf{C}^\dagger(\mathbf{k}).
\end{equation}
Together, these result in the stationary condition for minimizing the band SCFMI wavefunction
\begin{equation}
\mathbf 0
=
\left[
\left(\mathbf I - \mathbf S(\mathbf{k}) \mathbf P(\mathbf{k})\right)
\mathbf F(\mathbf{k})
\mathbf C(\mathbf k)
\mathbf{\boldsymbol{\sigma}}(\mathbf{k})^{-1}
\right]_{XX}.
\end{equation}

\subsection{Solid-state ALMO-EDA for molecular fragments\label{si_sec:cohesive_eda_theory}}

Solid-state ALMO-EDA for molecular fragments is designed for cases where we do not want to allow charge transfer between a fragment and its periodic image at the SCFMI level, such as in molecular crystals. 
While this would be possible to study using $\Gamma$-point ALMO-EDA with a large supercell, our solid-state ALMO-EDA allows us to take advantage of the translational symmetry of the problem.
This enables efficient convergence to the thermodynamic limit and allows the use of more accurate computational tools than would otherwise be too expensive, such as hybrid-DFT or all-electron calculations. 

In the language of the ALMO constraint, the orbitals are block-diagonal without
allowing mixing between periodic images
\begin{equation}
\mathbf C_\text{occ}^\text{molecular} =
\begin{pmatrix}
\mathbf C_{A_{R_1}} & \mathbf 0 & \mathbf 0 & \mathbf 0\\
\mathbf 0 & \mathbf C_{A_{R_2}}& \mathbf 0 & \mathbf 0\\
\mathbf 0 & \mathbf 0 & \mathbf C_{B_{R_1}}& \mathbf 0\\
\mathbf 0 & \mathbf 0 & \mathbf 0 & \mathbf C_{B_{R_2}}\\
\end{pmatrix},
\end{equation}
where $\{R_i\}$ indexes the corresponding Born-von Karman cell.
Further recognizing that the translational symmetry of the system now requires that 
$C_{A_{R_1}} = C_{A_{R_2}} \equiv C_{A}$, the momentum space ALMO constraint for {SCFMI with molecular fragments} is
\begin{equation}
\mathbf C_\text{occ}^\text{molecular}(\mathbf{k}) =
\delta_\mathbf{k ,\Gamma}
\begin{pmatrix}
\mathbf C_{A} & \mathbf 0\\
\mathbf 0 &  \mathbf C_{B}\\
\end{pmatrix}.
\end{equation}
Here, the frozen energy is computed using gas phase isolated fragments to generate $C_X$.
This results in an energy functional where we have orbitals defined purely for the $\mathbf{\Gamma}$-point parameterizing the Fock operator over all of k-space. 

Similar to before, 
the occupied MO metric tensor is given by 
\begin{equation}
\mathbf{\sigma}(\mathbf{k})
=
\mathbf{C}^\dagger\mathbf{S}(\mathbf{k})\mathbf{C},
\end{equation}
and the SCFMI 1-RDM is given by 
\begin{equation}
    \mathbf{P}(\mathbf{k}) = 
    \mathbf{C}
    \mathbf{\sigma}^{-1}(\mathbf{k})
    \mathbf{C}^\dagger.
\end{equation}
The stationary condition is likewise given by
\begin{align} 
    0
    &=
    -\frac{2}{N_k}
    \sum_k \text{Re}\left[
    \mathbf{\sigma}_\mathbf{k}^{-1}
    \mathbf{C}_\text{occ}^\dagger
    \mathbf{F_k}
    \left(
    \mathbf{I} - \mathbf{P_k S_k}
    \right)
    \mathbf{C}_\text{vir}
    \right].
\end{align}

As a technical note, for solid-state EDA with bulk fragments case, we took the virtual and occupied Fock matrices to be diagonal as a part of the pseudo-canonicalization step for GDM optimization.
In solid-state EDA with molecular fragments, we cannot assume that the Fock matrix is diagonal at all k.
Here, we choose to diagonalize only at ${\Gamma}$, so our occupied-occupied condition is
\begin{equation}
    \left[
    \mathbf{\sigma}_{\mathbf{\Gamma}}^{-1}
    \mathbf{C}_\text{occ}^{\dagger}
    \mathbf{F}_{\mathbf{\Gamma}}
    \mathbf{C}_\text{occ}
    \mathbf{\sigma}_{\mathbf{\Gamma}}^{-1}
    -\epsilon
    \right]_{XX} = 0.
\end{equation}

\section{Supplemental Data\label{si_sec:calcs}}

\subsection{Validation (Noble Gas Model)\label{si_sec:noble_gas}}

\begin{table}[ht]
\begin{ruledtabular}
\begin{tabular}{llllll}
\shortstack[l]{Lattice\\Constant (\AA)} &
\shortstack[l]{k-mesh or \\ supercell} &
$E_\text{supercell}^\text{(molecular)}$ (a.u) & 
$E_\text{kpts}^\text{(molecular)}$ (a.u) & 
$E_\text{kpts}^{\text{(bulk)}}$ (a.u) &
$E_\text{supercell}^\text{(bulk)}$ (a.u) \\ 
\hline
 $5.00$ & $1 \times 1 \times 1$ & $-257.6638917964$ & $-257.6638917964$ & $-257.6638917964$ & $-257.6638917964$\\
 $5.00$ & $2 \times 2 \times 2$ & $-257.7149786734$ & $-257.7149786746$ & $-257.7151358083$ & $-257.7151358070$\\
 $5.00$ & $3 \times 3 \times 3$ & $-257.7151828125$ & $-257.7151828120$ & $-257.7154277996$ & $-257.7154277998$\\
 $6.00$ & $1 \times 1 \times 1$ & $-257.7079354026$ & $-257.7079354026$ & $-257.7079354026$ & $-257.7079354026$\\
 $6.00$ & $2 \times 2 \times 2$ & $-257.7212676177$ & $-257.7212676179$ & $-257.7213172583$ & $-257.7213172580$\\
 $6.00$ & $3 \times 3 \times 3$ & $-257.7212939736$ & $-257.7212939742$ & $-257.7213512537$ & $-257.7213512524$\\
 $7.00$ & $1 \times 1 \times 1$ & $-257.7169739913$ & $-257.7169739913$ & $-257.7169739913$ & $-257.7169739913$\\
 $7.00$ & $2 \times 2 \times 2$ & $-257.7203895395$ & $-257.7203895396$ & $-257.7204099593$ & $-257.7204099591$\\
 $7.00$ & $3 \times 3 \times 3$ & $-257.7203925161$ & $-257.7203925162$ & $-257.7204203840$ & $-257.7204203838$\\
\end{tabular}
\end{ruledtabular}
\caption{\label{tab:cohesive_comp_total_energy}
\textbf{
Comparison of Ne$_2$ in a NaCl-type lattice across a range of different lattice constants and supercell/$\mathbf{k}$-mesh sizes.
}
}
\end{table}

We consider Ne$_2$ in a NaCl-type lattice as a simple toy model to validate the implementation of our 
solid-state ALMO-EDA against equivalent supercell approaches.
We chose this small system so that we can easily test larger supercells with a large all-electron basis set. 
These calculations applied the def2-TZVPD basis set with the PBE functional.
Across a range of lattice constants and supercell sizes, the results agree to  $\approx 10^{-9}$ a.u. indicating numerical equivalence between
the supercell and k-space methods.

\begin{figure}[tbh]
    \centering
    \includegraphics[width=1\linewidth]{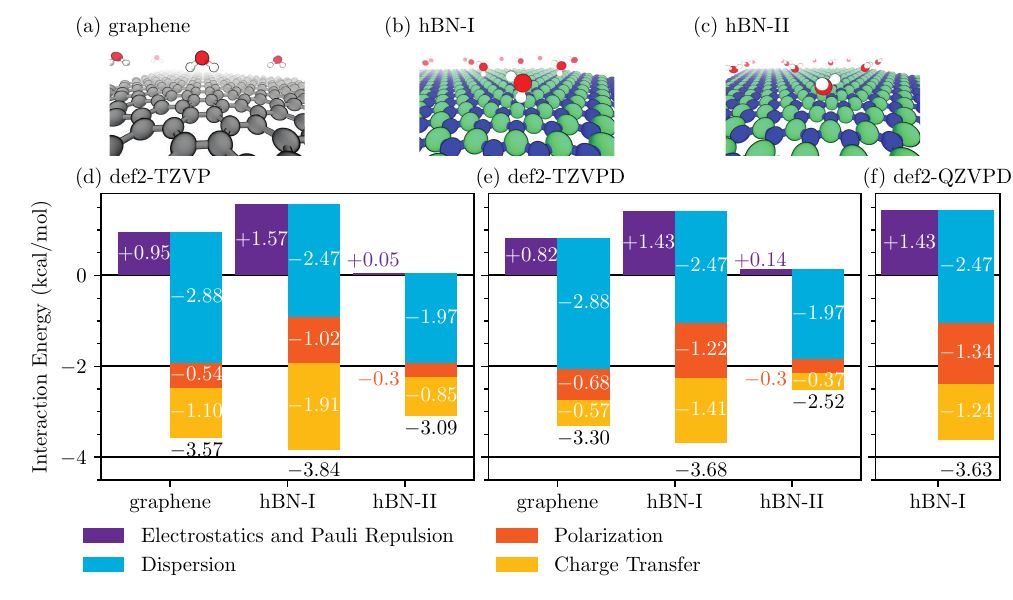}
    \caption{\textbf{Water adsorption on monolayer surfaces with several basis sets}. (a)-(c) Diagrams of the three systems studied. (d) EDA with def2-TZVP basis set. (e) EDA with def2-TZVPD basis set.
    (f) EDA with def2-QZVPD basis set}
    \label{SI_fig:water_adsorption}
\end{figure}

\subsection{Basis Set Convergence (Water Adsorption)\label{si_sec:water_surface}}

ALMO-EDA, as implemented in this paper, is known to lack a well-defined basis-set limit. 
As the basis-set size increases the more diffuse functions will contaminate the polarization energy with charge transfer character.
Applications of this type of EDA must therefore be careful not to use a basis set that is too large.
It is important that the relevant chemistry be captured, but that the basis remain compact.
Surface-adsorbate problems are cases where we would expect this contamination to be significant due to the unconfined nature of the adsorbates.
Additionally, this presents an opportunity to demonstrate our EDA method alongside many previous solid-state EDA applications.
Here we explore water adsorption onto graphene and hexagonal boron nitride (hBN), a canonical surface adsorption problem.
We compute the ALMO-EDA with the following basis sets: def2-TZVP, def2-TZVPD, and, on a single system, def2-QZVPD.
In this case, def2-TZVPD is a good balance between accuracy and charge-transfer contamination, so we are confident of our use of triple-$\zeta$ quality basis sets throughout the main text.

In \cref{SI_fig:water_adsorption} the EDA results are summarized for two configurations of water on hBN and one on graphene.
As one would expect, the hydrogen-bonding alignment on graphene and hBN-I yields larger polarization and charge-transfer contributions.

Examining the basis-set dependence explicitly, the total energy appears to be reasonably converged by the def2-TZVPD level, but the charge-transfer decreases uniformly as the basis-set size increases. There are two major contributors to this change.

First, the counterpoise correction decreases as the basis set approaches completeness.
For graphene, hBN-I and hBN-II, the counterpoise corrections decrease from $1.2$, $1.4$, and $0.6$ kcal/mol with def2-TZVP, respectively, to a consistent $0.1$ kcal/mol with def2-TZVPD.
For hBN-I, the counterpoise correction further reduces to $0.005$ kcal/mol with def2-QZVPD.
The large counterpoise corrections with def2-TZVP indicate that
the basis is missing important aspects of the fragments.
Including diffuse functions with def2-TZVPD largely eliminates the basis-set superposition error and captures the interaction energy well.
Further expanding the basis set to def2-QZVPD does not notably alter the frozen or total interaction energies.

Another contributor to the decrease in charge transfer is contamination with polarization.
While we cannot separate contamination from genuine polarization increases due to a better fragment wavefunction, the large change in the counterpoise correction between def2-TZVP and def2-TZVPD suggests that additional flexibility is important for correctly describing polarization in these systems. 
In all cases, the qualitative picture does not depend on the basis.
So, a triple-$\zeta$ basis should strike the right balance between charge transfer contamination and completeness.

For these calculations we applied PBE \cite{perdew_generalized_1996} in combination with the D3(0) dispersion correction\cite{grimme_consistent_2010} in an all-electron calculation employing the def2-TZVP, def2-TZVPD and def2-QZVPD basis sets and the rij-def2-TZVP auxiliary basis.\cite{weigend_balanced_2005, weigend_ri-mp2_1998, pritchard_new_2019}
We used a $10\times10\times1$ k-mesh for all systems.
For geometry optimizations, we used Quantum Espresso, with PBE, PAW pseudopotentials, and a $2\times2\times1$ k-mesh at 60 Ry. cutoff, and the D3 dispersion correction.

\subsection{Molecular Crystals: Benzene\label{si_sec:benzene}}
Benzene is a valuable benchmark for our solid-state ALMO-EDA scheme because of its ubiquity.
\cite{wen_practical_2012, yang_ab_2014, moellmann_dft-d3_2014}
Benzene is largely held together by dispersion interactions, and
the other energetic contributions are nearly negligible. Of the total 
$-13.3$ 
kcal/mol cohesive energy, $-12.7$ kcal/mol is from dispersion, while 
small polarization and charge transfer contributions of $-0.6$ and $-2.1$ kcal/mol, respectively, are nearly entirely counteracted by the electrostatic and Pauli repulsion term of 
$+2.2$ kcal/mol.
The overall cohesive energy, while not the focus of this study, is in 
agreement with benchmark results.\cite{moellmann_dft-d3_2014}
The agreement between our intuition and the existing literature provides confidence in the utility of ALMO-EDA for decomposing cohesive energies.

\subsection{Molecular Crystals: Dimer EDA Calculations\label{si_sec:molecular_eda}}

To better understand how much of the cohesive energy EDA of the molecular crystals is due to the local environment as opposed to the longer-range crystal environment, we calculate the molecular EDA of the four dimers.
The results of this analysis are presented in \cref{SI_tab:dimer_EDA} and discussed in the main text.
We note that the dispersion interaction of the solid is the least well-represented by the dimer models, which aligns with intuitive understandings of the energy components.
In the case of aspirin I, the dimer model only accounts for $-3.6$ kcal/mol of the $-21.2$ kcal/mol with solid-state ALMO-EDA.

The dimer calculations were run with a development version of Q-Chem 6.4.\cite{epifanovsky_software_2021}
Dimer geometries were extracted from the optimized molecular crystal structures.
As in the periodic case, the def2-TZVP basis was applied with the PBE functional.\cite{weigend_balanced_2005, perdew_generalized_1996}
The AO-based polarization was used for the EDA to ensure comparability with our solid-state ALMO-EDA results. 

\begin{table}[ht]
    \begin{ruledtabular}
        \begin{tabular}{lrrrr}
Molecular Crystal & 
$E_\text{disp}$ (kcal/mol)  &
$E_\text{elec. + Pauli}$ (kcal/mol)&
$E_\text{pol}$ (kcal/mol)&
$E_\text{ct}$ (kcal/mol)\\
        \hline
    Aspirin I dimer &
    $-3.6$ & $8.7$ & $-7.1$ & $-10.9$
\\ Aspirin IV dimer  &      
    $-3.8$ & $10.2$ & $-7.5$ & $-11.5$
\\ Aspirin-Theophylline dimer &
    $-3.7$ & $4.5$ & $-5.0$ & $-7.3$
\\ Aspirin-Carbamazepine dimer &
    $-3.4$ & $5.9$ & $-5.7$ & $-8.6$
        \end{tabular}
    \end{ruledtabular}
    \caption{\textbf{Gas-phase EDA of the central molecular crystal dimers}}
    \label{SI_tab:dimer_EDA}
\end{table}

\subsection{Molecular Crystals: Density-Fitting and Basis Set Superposition Corrections\label{si_sec:df_errors}}

The incompleteness of the main basis set is responsible for the standard basis set superposition error (BSSE), while the incompleteness of the auxiliary basis is responsible for density fitting errors. 
In this section, we show the correction schemes applied to ameliorate these errors.

The BSSE correction of Boys and Bernardi\cite{boys_calculation_1970} was included for each subsystem by including `ghost' atomic centers within 4~\AA \ of the active subsystem such that the subsystems contained 84 centers (benzene),  117 centers (aspirin I), 126 centers   (aspirin IV), 126 centers (aspirin theophylline), and 115 or 150 centers (aspirin carbamazepine for the aspirin and carbamazepine subsystems respectively).
\Cref{tab:df_error} shows the value of the counterpoise correction for each of the molecular crystal systems.
It is important to note, the correction is $\approx 10-15\%$ of the uncorrected charge transfer, so it does not qualitatively change the results. 

We additionally ran a benchmark calculation for the convergence of our BSSE distance cutoff. 
With a cutoff of 5~\AA \, the charge transfer energies of benzene and aspirin IV increased by 0.1 and 0.2  kcal/mol from the 4~\AA \ cutoff results, respectively, indicating good convergence.

The all-electron calculations presented for the molecular crystals are made possible with density fitting (DF) DFT, which
substantially speeds up the evaluation of the two-electron integrals.\cite{dinh_efficient_2026}
Our molecular crystal calculations used density fitting to compute frozen, polarization, and charge-transfer effects, but not for isolated fragments.
This discrepancy introduces an extra energetic term to the frozen energy, which we can approximately correct by subtracting the difference between fragments computed with 
density fitting and without density fitting.
This density fitting error for the fragments is shown in \cref{tab:df_error}.

Another correction we consider is the auxiliary basis-set superposition error (ABSSE) as defined by Thirman and Head-Gordon.\cite{thirman_energy_2015}
Because diffuse auxiliary functions can improve the approximate integrals on different fragments, the frozen step will additionally include this energetic difference in addition to the physical energetic changes.
We can correct for this difference via calculations including ghost auxiliary functions.\cite{thirman_energy_2015}

Together, these two terms result in a frozen energy correction of 
\begin{equation}
    E_\text{frz} = E_\text{frz}^\text{tot} - E_\text{frgm}^\text{tot} + E_\text{DF err} + E_\text{ABSSE}. 
\end{equation}
Here, $E_\text{frz}^\text{tot}$ refers to the absolute energy of the frozen wavefunction while $E_\text{frz}$ refers to the energy difference as in the main text.
In Ref.~\citenum{thirman_energy_2015} they additionally remove the ABSSE contribution from the charge transfer term to avoid double counting, but here such double counting is avoided because the BSSE correction is computed without density fitting.
We note that this correction is a part of the electrostatic and Pauli repulsion term, and represents an $\approx 5\%$ correction to the uncorrected electrostatic and Pauli repulsion contribution.

\begin{table}[ht]
    \begin{ruledtabular}
        \begin{tabular}{llll}
Molecular Crystal & BSSE Correction (kcal/mol) & DF 
Error (kcal/mol) & ABSSE Correction (kcal/mol)\\
\hline
Benzene & $0.73$ & $0.15$ & $-0.05$\\
Aspirin I & $1.91$ & $0.25$ & $-0.06$\\
Aspirin IV & $1.86$ & $0.25$ & $-0.06$\\
Aspirin Theophylline & $1.94$ & $0.27$ & $-0.06$\\
Aspirin Carbamazepine & $1.80$ & $0.31$ & $-0.07$\\
        \end{tabular}
    \end{ruledtabular}
    \caption{\textbf{Energetic corrections for the set of molecular crystals.}}
    \label{tab:df_error}
\end{table}

\subsection{\texorpdfstring{MoS$_2$/WSe$_2$}{MoS2WSe2}: Convergence\label{si_sec:MoS2WSe2_convergence}}

Here, we demonstrate the convergence behavior of the band gaps of MoS$_2$/WSe$_2$ with respect to the k-point sampling. 
For this example, we use the $AB_\text{Se}$ stacking as our test system. 

~\cref{SI_tab:MoS2WSe2_converge} demonstrates how the direct band gap (at the $K$ point in the Brillouin zone), changes as the k-space sampling approaches the thermodynamic limit.
Each of these calculations used a Monkhorst-Pack mesh centered on the $K$-point.
We note that these results do not contain the strain or spin-orbit correction included in the main text.
The data indicate that increasing the k-mesh size largely corrects the fragment contribution to the band gap while the other contributions are well-described even at a small mesh size.
Between $4\times4\times1$ and $7\times7\times1$, the fragment band gap decreases from $0.60$ eV to $0.53$ eV while the difference between the charge transfer band gap and the fragment band gap decreases from $0.43$ eV to $0.42$ eV.
This demonstrates that our use of $5\times5\times1$ for band structures and $7\times7\times1$ for energies and band gaps is sufficiently converged.

\begin{table}[ht]
\begin{ruledtabular}
    \begin{tabular}{lrrrr}
        EDA term & 
        $4\times4\times1$ & 
        $5\times5\times1$ & 
        $6\times6\times1$ & 
        $7\times7\times1$ 
         \\
        \hline
        Fragments ($K\to  K$) 
        &0.60 eV & 0.55 eV & 0.54 eV &  0.53 eV  \\
        Frozen  ($K\to  K$)  
        &0.75 eV & 0.70 eV & 0.68 eV &  0.68 eV \\
        Polarization  ($K\to  K$)  
        & 0.88 eV & 0.82 eV & 0.81 eV & 0.80 eV \\
        Charge Transfer  ($K\to  K$)  
        & 1.03 eV & 0.97 eV & 0.95 eV & 0.95 eV \\
        \end{tabular}
    \end{ruledtabular}
    \caption{
    \textbf{Convergence of the band gaps for MoS$_2$/WSe$_2$  with increasing k-mesh density.}
    }
    \label{SI_tab:MoS2WSe2_converge}
\end{table}

\subsection{\texorpdfstring{MoS$_2$/WSe$_2$}{MoS2WSe2}: Spin-Orbit Coupling and Lattice Strain\label{si_sec:MoS2WSe2_SOC_and_strain}}

\begin{figure}[htb]
    \centering
    \includegraphics[width=1.\linewidth]{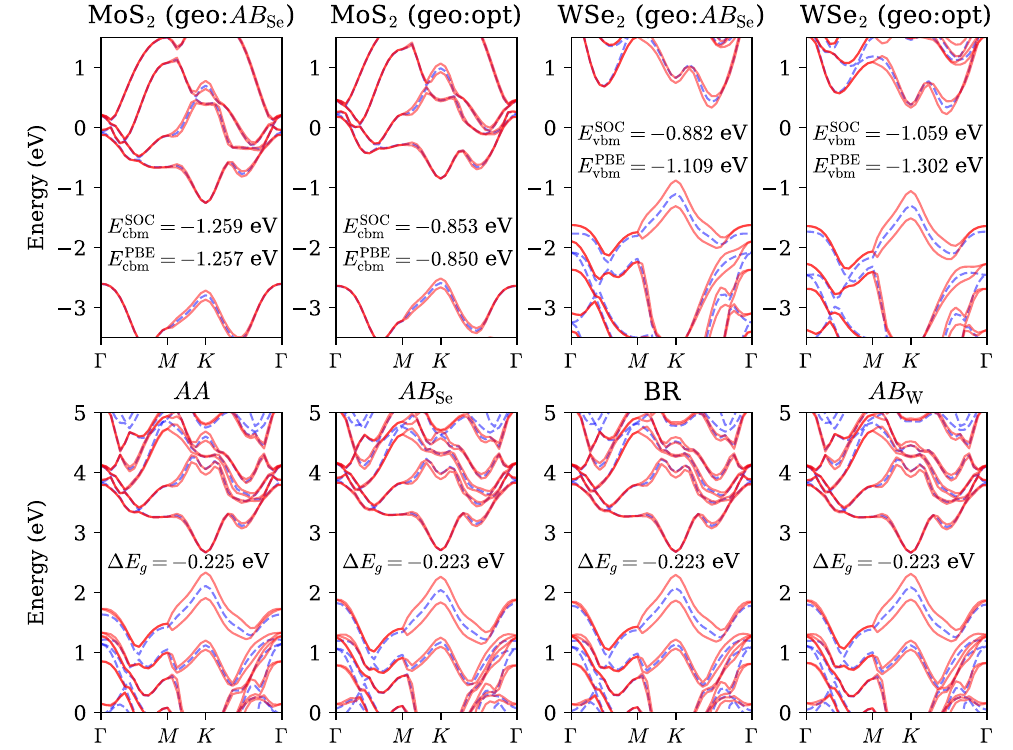}
    \caption{\textbf{Effect of spin-orbit coupling and lattice strain on MoS$_2$/WSe$_2$ bilayer}
    Dashed blue lines correspond to PBE without spin-orbit coupling. 
    Solid red lines correspond to PBE with spin-orbit coupling.
    Top: Band structures of the two monolayers in the optimized $AB_\text{Se}$ geometry and their optimized isolated geometries. The annotated values indicate the value of the conduction/valence band minimum/maximum (cbm/vbm) with and without spin-orbit coupling.
    Bottom:
    Band structures of the four stacking arrangements of MoS$_2$/WSe$_2$.
    The annotated values indicate the differences in band gaps between calculations with and without spin-orbit coupling.
    }
    \label{SI_fig:soc_moire}
\end{figure}

It is well-understood that spin-orbit coupling (SOC) has a large renormalization effect on the band structures of heavy-element-containing compounds.
In \cref{SI_fig:soc_moire}, we demonstrate that the SOC effects are almost entirely from the WSe$_2$ layer.
The SOC splitting of the valence band in WSe$_2$ leads to a band gap decrease of 
$-0.22$ to $-0.23$ eV. 
In the optimized $AB_\text{Se}$ geometry and the optimized isolated-layer geometry, SOC increases the VBM of WSe$_2$ by $0.227$ and $0.243$ eV, respectively.
Meanwhile, the CBM of MoS$_2$ is nearly unchanged by inclusion of SOC in either geometry.
This indicates that neither stacking nor strain has a meaningful effect on the SOC, and we can include band gap shifts ($\Delta E_g$) from \cref{SI_fig:soc_moire} as constant shifts to our EDA band gaps.

Representing the moir{\'e} lattice as perfectly periodic stacking arrangements introduces strain into the lattice which would not be present in the true moir{\'e} lattice. 
In the periodic stackings, MoS$_2$ is slightly stretched and WSe$_2$ is slightly compressed.
In Ref. \citenum{zhang_interlayer_2017} they correct for this strain in their band gaps via comparison with the optimized isolated monolayer bands.
In \cref{SI_fig:soc_moire}, the CBM of MoS$_2$ rises upon relaxation while the VBM of WSe$_2$ decreases upon relaxation.
Together, these account for an increase in the bilayer band gap of $0.58$ and $0.60$ eV with and without SOC, respectively. 

Combining the SOC and the strain, we expect a band-gap increase of about $0.4$ eV in our hybrid calculations without SOC. 
For the correction term presented in the main text, the SOC contribution is from \cref{SI_fig:soc_moire} while the strain contribution is calculated for each system using the hybrid-DFT method described in the main text.

For the calculations in this section, we used Quantum Espresso\cite{epifanovsky_software_2021}, with PBE\cite{perdew_generalized_1996}, and PAW pseudopotentials. The single-point calculations for bands used a 120 Ry. cutoff energy.

\subsection{Perovskite Heterostructures: Spin-Orbit Coupling\label{si_ssec:perovskite_soc}}

\begin{figure}[ht]
    \centering
    \includegraphics[width=\linewidth]{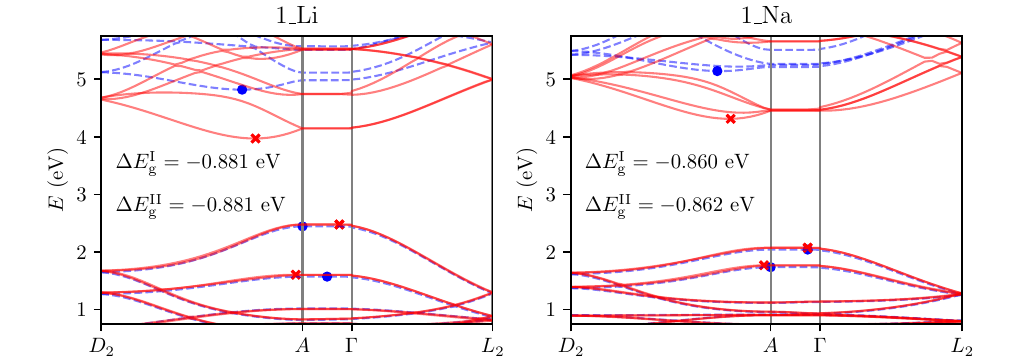}
    \caption{\textbf{Effect of spin-orbit coupling and lattice strain on perovskite heterostructures}
    Dashed blue lines correspond to PBE without spin-orbit coupling. 
    Solid red lines correspond to PBE with spin-orbit coupling.
    The annotated values indicate the difference in type-I and type-II band gaps between calculations with and without spin-orbit coupling.
    {
    The band minima and maxima along this path are denoted with blue circle markers (no spin-orbit coupling) and red x markers (including spin-orbit coupling)
    }
    }
    \label{SI_fig:perovskite_SOC}
\end{figure}

Because the perovskite heterostructures contain lead, we expect that spin-orbit effects are significant. 
In Fig.~\ref{SI_fig:perovskite_SOC}, we show that inclusion of SOC is nearly constant across both systems.
{
The k-path is extracted from the suggested path in Ref.~\citenum{hinuma_band_2017}, following Ref.~\citenum{deshmukh_tuning_2025}.
We note that the conduction band minima and valence band maxima do not lie along this k-path; however, the error introduced by constructing the SOC correction from this path is small.
For instance, the path $\Gamma-[-\frac12,0,\frac12]$ yielded a smaller gap and a change in the SOC corrections of $> 0.03$ eV.

For these calculations, we follow the band assignments of Fig. 4 in Ref.~\citenum{deshmukh_tuning_2025}, and therefore account for the subsystem band crossings when finding our valence band maxima.
}

The conduction bands are lowered by the SOC while the valence bands are largely unchanged, so both the type-I and type-II gap in each system are shifted by nearly the same amount. 
In the main text, the SOC correction is taken from Fig.~\ref{SI_fig:perovskite_SOC}.

For the calculations in this section, we used Quantum Espresso\cite{epifanovsky_software_2021}, with PBE\cite{perdew_generalized_1996}, and PAW pseudopotentials. The single-point calculations for bands used a 70 Ry. cutoff energy.

\subsection{Perovskite Heterostructures: Mulliken Population\label{si_ssec:perovskite_mulliken}}

At the charge transfer step, assigning bands to a specific subsystem is not uniquely defined. 
Here, we use a Mulliken population scheme averaged over the full Brillouin zone for each occupied and virtual band individually. 
Tab.~\ref{si_tab:mulliken_perovskite} shows the normalized results of this population scheme with subsystem contributions being defined by the sum of their constituent atoms' contributions. 

For 1\_Na, the CBM contains 29\% \ce{PbCl2} character, but the \ce{PbCl4^{2-}} character is predominate at 63\%.
Generally, the VBM and CBM of both systems are well-characterized by a single subsystem, so we are confident in the charge transfer contributions in type-II 1\_Li and type-I 1\_Na. 

The lower lying bands are less clear cut because there is genuine character mixing, and also because averaging the subsystem character across the whole Brillouin zone ignores band crossings between subsystem bands. 
In the main text, we base the charge transfer analysis off of the first valence or conduction band (measured from the Fermi level) with more than 15\% character of a subsystem. 
Increasing this cutoff would eventually increase the charge transfer contribution in the band gap analyses for the type-I 1\_Li and type-II 1\_Na since we cannot determine those gaps without decomposing the non-frontier bands with some form of population analysis.

\begin{table}[ht]
    \begin{ruledtabular}
        \begin{tabular}{clllll|llll}
 & Subsystem       & VBM-3& VBM-2& VBM-1& VBM  & CBM  & CBM+1& CBM+2& CBM+3\\
\hline
1\_Li & \ce{PbCl4^{2-}} & 0.39 & 0.38 & \textbf{0.58}& 0.00 & \textbf{0.99} & 0.80 & 0.26 & 0.28 \\
1\_Li & \ce{PbCl2}      & 0.45 & 0.43 & 0.28 & \textbf{0.77} & 0.00 & 0.09 & \textbf{0.29} & 0.19 \\
1\_Li & other subsystems & 0.17 & 0.19 & 0.14 & 0.23 & 0.01 & 0.11 & 0.45 & 0.53 \\
 \hline
1\_Na & \ce{PbCl4^{2-}} & 0.53 & 0.18 & 0.28 & \textbf{0.96} & \textbf{0.63} & 0.17 & 0.25 & 0.51 \\
1\_Na & \ce{PbCl2}      & 0.12 & 0.16 & \textbf{0.33} & 0.01 & \textbf{0.29} & 0.69 & 0.49 & 0.21 \\
1\_Na & other subsystems & 0.35 & 0.65 & 0.40 & 0.03 & 0.09 & 0.15 & 0.26 & 0.29 \\
        \end{tabular}
    \end{ruledtabular}
    \caption{\textbf{Normalized Mulliken population analysis for the perovksite heterostructure bands} The bold numbers indicate the valence and conduction bands presented in the main text.}
    \label{si_tab:mulliken_perovskite}
\end{table}

\end{document}